**Virus Transmission Risk in Urban Rail Systems: A Microscopic Simulation-based Analysis of Spatio-temporal Characteristics**

**Jiali Zhou, Corresponding Author**
Department of Civil and Environmental Engineering
Northeastern University, Boston MA 02115
Tel: 412-708-2493; Email: zhou.jiali1@northeastern.edu

**Haris N. Koutsopoulos**
Department of Civil and Environmental Engineering
Northeastern University, Boston MA 02115
Tel: 617-373-6263; Email: h.koutsopoulos@northeastern.edu

**ABSTRACT**
Transmission risk of air-borne diseases in public transportation systems is a concern. The paper proposes a modified Wells-Riley model for risk analysis in public transportation systems to capture the passenger flow characteristics, including spatial and temporal patterns in terms of number of boarding, alighting passengers, and number of infectors. The model is utilized to assess overall risk as a function of OD flows, actual operations, and factors such as mask wearing, and ventilation. The model is integrated with a microscopic simulation model of subway operations (SimMETRO). Using actual data from a subway system, a case study explores the impact of different factors on transmission risk, including mask-wearing, ventilation rates, infectiousness levels of disease and carrier rates. In general, mask-wearing and ventilation are effective under various demand levels, infectiousness levels, and carrier rates. Mask-wearing is more effective in mitigating risks. Impacts from operations and service frequency are also evaluated, emphasizing the importance of maintaining reliable, frequent operations in lowering transmission risks. Risk spatial patterns are also explored, highlighting locations of higher risk.

## INTRODUCTION

COVID-19, an infectious disease caused by the virus SARS-CoV-2 [1] has greatly impacted the lives of people at global scale. In light of the infectiousness of COVID-19, social distancing, mask-wearing, and contact tracing are promoted as important means to contain the spread of the disease. Public transportation moving, large amount of people in urban areas ,is also part of the discussion.

Various studies in the literature have looked into the role of public transportation in the transmission process. Studies have also looked at the impact of pandemics on public transportation ridership, attitudes towards using public transportation, etc. New York's Metropolitan Transportation Authority (MTA) reported a subway ridership decline more than 87% and a bus ridership decline of 60% by mid-April 2020 due to COVID-19 [2]. In response to the decrease in demand, many transit agencies reduced service. The MTA for example, cut its services by 25% in light of the decline in ridership [3]. The Massachusetts Bay Transportation Authority (MBTA) reported more than 90% ridership decline in April 2020 [4].

Researchers have studied the role of public transportation systems in the transmission of airborne diseases. Goscé et al.'s study [5] showed a correlation between the use of public transport and the spread of influenza-like illnesses (ILI) in the London underground after comparing Automatic Fare Collection (AFC) data and ILI data. The ILI data was collected by the UK National Health Service (NHS) in London boroughs on a daily basis, covering more than 40% of the population.

Harris [6] studied the MTA subway lines in the New York City area and concluded that the subway system could be a major disseminator for COVID-19, pointing out that the drop in COVID-19 cases coincided with the decrease in ridership in the subway system. However, researchers question the conclusions of the paper, arguing that other activities, including school and office activities, dropped at the same time. Zheng et al. [7] study the spatial pattern of COVD-19 transmission via public and private transportation in China and found a significant association between the frequency of flights, trains, and buses from Wuhan, China and daily cases in other cities in China.

Regardless of the arguments regarding the correlation between the COVID-19 spread and public transportation use, stations and transit vehicles are indoor environments, often densely crowded with unknown ventilation, and hence present risks of transmission of airborne infections. Nasir et al. [8] argue that the most important factors influencing airborne disease transmission in transport environments, including stations and vehicles, are related to the ventilation and HVAC systems, space configurations, hygiene maintenance, and access/entryway control. The National Air Filtration Association (NAFA) has also emphasized and evaluated the transmission risk of air-borne diseases, such as Influenza, Severe Acute Respiratory Syndrome (SARS) and Tuberculosis etc. in indoor spaces [9]. Luo et al. [10] report two outbreaks of COVID-19 transmission in bus trips and discuss the risk of transmission in buses.

Researchers have proposed models at different scales and resolutions to quantify risks and model the transmission process. Tuite et al. [11] applied a modified "susceptible-exposed-infectious-recovered" (SEIR) framework to model the day-to-day COVID-19 transmission patterns, accounting for public health interventions and different severities of clinical symptoms in Ontario, Canada. Prem et al. [12] used an age-structured SEIR framework and contact matrices to model the transmission process under different control measures. Sustained distancing is important for containing the magnitude of the epidemic and sudden lifting of interventions can lead to secondary outbreaks.

These studies applied and adapted the contact network of the population and the popular "susceptible-infectious-recovered" framework for transmission modeling. However, such studies focus more on the population in larger areas or at a city-wide scale and consider the transmission risk over a time horizon of days or months. They model the risk level at a macroscopic level.

Other studies focus on the transmission risk of airborne diseases, aiming to assess the risks in small, confined spaces and shorter time periods, normally within a few hours, conditions that characterize most of urban transit trips. Wiley et al. [13] proposed a model, for an epidemiological study of a measles outbreak. The model was based on the concept of 'quantum of infection', proposed by Wells [14], which indicates the transmission capability of a certain disease. The model, known as the Wells-Riley model has been used extensively by researchers to study the transmission risk inside hospital wards, classrooms, offices, and transit vehicles for various respiratory diseases, including measles, influenza, SARS, etc. [9, 15]. The original model estimates the expected number of infections as a function of variables for ventilation rate, exposure time, and number of carriers. It has also been used to study the effectiveness of methods to mitigate the risks of transmission in confined spaces, such as increasing ventilation capacity [9].

Other researchers developed a variation of the original Wells-Riley model to account for the impact of mask-wearing. Fennelly et al. [16] used a mask-wearing-specific Wells-Riley model to conclude that mask-wearing can be useful for lowering the risk of transmission. Dai et al. [17] applied the Wells-Riley model to study the transmission risk of COVID-19 under various ventilation levels and in different indoor environments, including aircraft cabins, classrooms, offices, and buses. They concluded that the risk of airborne transmission in these indoor environments is not negligible.

Some researchers used the Wells-Riley model to explore the transmission risk in transportation. Ko et al. [18], Chen et al. [19], and Furuya [20] applied the model to assess the risk in commercial airline and train trips. Andrews et al. [21] conducted a complete risk analysis for various means of public transportation, including bus, rail, taxi, etc. and concluded that tuberculosis transmission may occur in South Africa's public transportation.

The various studies above, although used the Wells-Riley model for risk analysis of indoor environments, they treated the public transportation vehicle as a close-to-static indoor environment, similar to classrooms or offices, with a fixed number of people and infectors inside. The exposure time is also assumed to be constant.

However, urban public transportation vehicles travel through different areas and keep picking up and dropping off passengers in a more dynamic manner. The number of people on board varies spatially and temporally from section to section. Previous studies do not capture the dynamic and spatial characteristics of public transportation demand.

This paper aims to model the risk of transmission of airborne diseases in public transportation systems using the Wells-Riley equation and a microscopic urban rail simulation platform. The model takes into account the spatio-temporal characteristics of the demand using the system, service delivery characteristics, as well as mitigation measures related to passenger mask-wearing behavior and vehicle ventilation performance.

The paper is organized as follows. Section 2 discusses the derivation of a Wells-Riley based model incorporating the demand and supply dynamics in public transportation systems. Section 3 presents a case study that evaluates the effectiveness of different mitigation strategies for various demand levels and disease characteristics. The case study also explores the impact of system operating characteristics (such as frequency and reliability) on the risk. Section 4 concludes the paper.

**METHODOLOGY**
**The Wells-Riley Model**

The Wells-Riley model was proposed by Riley et al. [13] to evaluate the probability of infection in indoor premises, considering the intake dose of airborne pathogens in terms of the number of quanta. The equation is shown below.

$$P = 1 - \exp\left(-\frac{Ipqt}{Q}\right) \quad (1)$$

where, $P$ is the probability of infection, $I$ the number of infectors (carriers), $p$ the breathing rate per person (m$^3$/hour), $q$ the quanta generation rate (1/hour), $t$ the exposure time length (hour), $Q$ the room ventilation rate of clean air (m$^3$/hour). It is worth noticing that $q$, the quanta generation rate (1/hour), is not a physical unit. Instead, it captures the number of infectious particles that are exhaled to the environment and also the infectivity of the particles. The value of $q$ can be back-calculated from epidemiological studies of outbreak cases using equation (1).

The Wells-Riley equation assumes a steady state of infectious particles in the environment and that the indoor air is well-mixed. The biological decay of the airborne pathogens is assumed negligible. The model has been used to evaluate risks of transmission in various indoor environments, including hospitals, office rooms, classrooms, transit vehicles, trains, and airplane cabins [13, 15, 16, 17, 18, 19, 20, 21].

The original model (equation (1)) has been modified in the literature to incorporate additional factors. For respiratory diseases, the use of respirator or mask can reduce the number of inhaled infectious particles. The term $(-\frac{Ipqt}{Q})$ in equation (1) is a measure of the intake dose. Fennelly et al. [16] developed a form of Wells-Riley equation considering mask-wearing to capture the impact of masks on the exhaled infectious dose. Similarly, Furuya [20] captures the filtration effect of respirators, including surgical mask, as inhaling protection, by adding a discounting factor towards $p$, the breathing rate per person (m³/hour).

In the case of both infectors and susceptible persons wearing a mask, the probability of infection can be expressed as:

$$P = 1 - \exp\left(-\frac{I(F_m \times p)(R_m \times q)t}{Q}\right) \qquad (2)$$

Wearing masks protects through filtering both inward (inhaling) and outward (exhaling) air movements. The infectious particles exhaled by infectors are filtered by a certain rate, $R_m$, the particle penetration rate of respirator, $R_m = 1$ represents a non-effective mask with all infectious particles from the infector exhaled into the air. $R_m = 0$ means 100% particle blockage. On the other hand, the particles inhaled by susceptible persons could be filtered by a certain rate, $F_m$ ($F_m = 1$ means completely ineffective mask and $F_m = 0$ 100% particle blockage).

The various studies into the transmission risk in public transportation vehicles, neglect the passenger flow dynamics in terms of their origin and destination stations. They assume a fixed number of people inside those areas for a given time period. This ignores passenger flow characteristics, including spatial and temporal patterns in terms of number of boarding, alighting passengers, and the number of infectors and their distribution along the line based on their Origin-Destination (OD) pair. The approach proposed in the next section aims at relaxing the assumptions of earlier studies.

**Modeling Infection Risk for Transit Operations**

We illustrate the approach using the following example (Figure 1) of a subway line with 4 stations. There are three passengers in the system. Passenger (pax) 1 is susceptible, while passengers 2 and 3 are infected (carriers). The train is moving from station 1 to station 4. Passenger 1 is traveling from station 2 to station 4, passenger 2 from station 1 to station 3, and passenger 3 from station 2 to station 4.

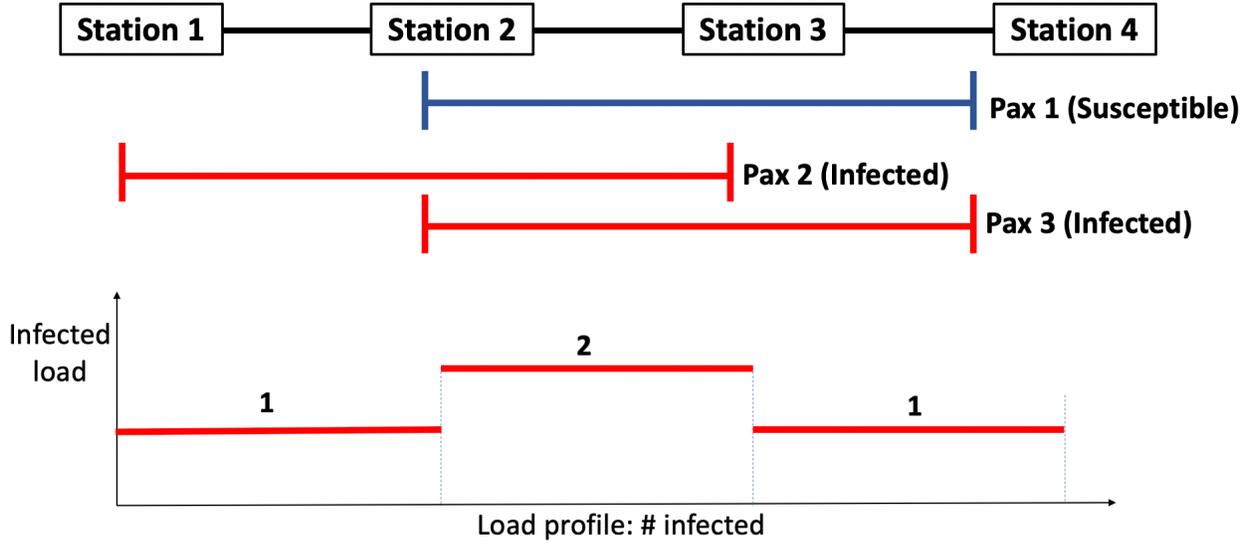

**Figure 1 Illustration for passenger transmission modeling**

Passenger 1's risk of being infected comes from their interactions with passengers 2 and 3 when they share their transit rides in the same vehicle. The probability of not getting infected by pax 2, based on equation (1) is:

$$P(\text{not infected by pax 2}) = 1 - \left(1 - \exp\left(-\frac{qpt_{23}}{Q}\right)\right) = \exp\left(-\frac{qpt_{23}}{Q}\right) \quad (3)$$

where $t_{23}$ is the time for the train to travel from station 2 to 3, which is the time pax 1 and pax 2 spend together in the same vehicle.

Similarly, pax 1's probability of not infected by pax 3 is:

$$P(\text{not infected by pax 3}) = 1 - \left(1 - \exp\left(-\frac{qpt_{24}}{Q}\right)\right) = \exp\left(-\frac{qpt_{24}}{Q}\right) \quad (4)$$

where $t_{24}$ is the time for the train to travel from station 2 to 4.

Thus, the overall probability of not being infected is:

$$P(1 \text{ not infected}) = P(\text{not infected by 2}) \times P(\text{not infected by 3})$$
$$= \exp\left(-\frac{qpt_{23}}{Q}\right) \times \exp\left(-\frac{qpt_{24}}{Q}\right) = \exp\left(-\frac{qp(t_{23}+t_{24})}{Q}\right) = \exp\left(-\frac{2qpt_{23}}{Q}\right) \times \exp\left(-\frac{qpt_{34}}{Q}\right) \quad (5)$$

Equation (5) indicates that the infection risk of a passenger is a function of the load of infected passengers onboard each segment of their trip. Hence, it can be easily generalized for any passenger traveling on OD pair *ij*, assuming that the number of infected passengers on any segment of the trip is known.

$$P(\text{not infected on OD } ij) = \prod_{s \in ij} \exp\left(-\frac{L_s qpt_s}{Q}\right) \quad (6)$$

Therefore, the probability of infected, $P_{ij}$, is given by:

$$P_{ij} = 1 - \prod_{s \in ij} \exp\left(-\frac{L_s q p t_s}{Q}\right) \tag{7}$$

where $P_{ij}$ is the probability of infection for a passenger traveling on OD pair $ij$, $L_s$ the number of infected carriers on segment s, $t_s$ travel time on segment s.

Given the probability $P_{ij}$, the distribution of the number of infections among passengers traveling on OD $ij$ follows the binomial distribution and the expected number of infections, $R_{ij}$, is

$$r_{ij} = P_{ij} D_{ij} \tag{8}$$

where $D_{ij}$ is the OD flow *from i to j* (susceptible).

The model can be easily extended to accommodate mask-wearing behavior. Let $f_m$ be the fraction of mask-wearing passengers, $f_i$ the fraction of infectors (carriers), and $N_s$ the number of boarding passengers at station $s$.

For each OD pair $ij$, passengers boarding train $k$ can be divided into 3 subgroups:
1. $D_{ijk}^{susceptible}$, number of susceptible passengers from station $i$ to station $j$ on train $k$;
2. $D_{ijk}^{infected,mask}$, number of infected passengers wearing a mask from station $i$ to station $j$ on train $k$;
3. $D_{ijk}^{infected,no\_mask}$, number of infected passengers not wearing a mask from station $i$ to station $j$ on train $k$.

The passenger load on segment $s$, between stations $s$ and $s+1$, can be calculated by subgroup as follows:

$$L_{s,k}^{susceptible} = \sum_{i=1}^{S} \sum_{j=s+1}^{S} D_{ijk}^{susceptible} \tag{9}$$

$$L_{s,k}^{infected,mask} = \sum_{i=1}^{S} \sum_{j=s+1}^{S} D_{ijk}^{infected,mask} \tag{10}$$

$$L_{s,k}^{infected,no\_mask} = \sum_{i=1}^{S} \sum_{j=s+1}^{S} D_{ijk}^{infected,no\_mask} \tag{11}$$

where $L_{s,k}^{susceptible}$ is the susceptible passenger load leaving station $s$ on train $k$, $L_{s,k}^{infected,mask}$ the masked infector passenger load leaving station $s$ on train $k$, and $L_{s,k}^{infected,no\_mask}$ the non-masked infector passenger load leaving station $s$ on train $k$.

The probability $\overline{P_{i,j,k}^{masked}}$ of a susceptible person traveling from $i$ to $j$ on train $k$ and wearing a mask not infected is calculated by combining equations (6), (7), (10) and (11).

$$\overline{P_{i,j,k}^{masked}} = \prod_{s \in (i,j)} \exp\left(-\frac{L_{s,k}^{infected,mask}(F_m \times p)(R_m \times q)t}{Q}\right) \prod_{s \in (i,j)} \exp\left(-\frac{L_{s,k}^{infected,no\_mask}(F_m \times p)(1 \times q)t}{Q}\right) \tag{12}$$

Similarly,

$$\overline{P_{i,j,k}^{non\_masked}} = \prod_{s\in(i,j)} \exp\left(-\frac{L_{s,k}^{infected,mask}(1\times p)(R_m\times q)t}{Q}\right) \prod_{s\in(i,j)} \exp\left(-\frac{L_{s,k}^{infected,no\_mask}(1\times p)(1\times q)t}{Q}\right) \quad (13)$$

where $\overline{P_{i,j,k}^{non\_masked}}$ is the probability that a non-masked susceptible passenger traveling from station $i$ to station $j$ on train $k$ is not infected. $R_m$ as stated earlier is the effectiveness of a mask.

Consequently, the probability of infection for a passenger traveling from $i$ to $j$ is given by:

$$P_{i,j,k}^{masked} = 1 - \overline{P_{i,j,k}^{masked}} \quad (14)$$

$$P_{i,j,k}^{non\_masked} = 1 - \overline{P_{i,j,k}^{non\_masked}} \quad (15)$$

Using equations (9), (14), and (15), the expected number of infected passengers traveling on OD pair $ij$ and train $k$ is given by:

$$r_{ijk} = P_{i,j,k}^{non\_masked}(1-f_m)D_{ijk}^{susceptible} + P_{i,j,k}^{masked} f_m D_{ijk}^{susceptible} \quad (16)$$

The expected number of infections, $r_k$, onboard train $k$ is :

$$r_k = \sum_i \sum_j r_{ijk} \quad (17)$$

The expected number of infections per trip at the OD level is:

$$r_{ij} = \frac{\sum_k r_{ijk}}{K} \quad (18)$$

where $K$ is the number of train trips.

The total expected number of infections $r$ at the system level is given by:

$$r = \sum_i \sum_j \sum_k r_{ijk} \quad (19)$$

Another metric of interest is the probability of infection taking into account (weighted by) the distribution of the trips in the system. For a random passenger using train $k$ the probability of infection $P_k$ is given by:

$$P_k = \frac{r_k}{N_k} \quad (20)$$

where $N_k$ is the total number of passengers on train $k$.
At the system level the overall probability of infection $P$ is:

$$P = \frac{r}{N} \quad (21)$$

where $N$ is the total number of passengers in the system.

These metrics, $r$ and $P$, are useful measures of risk and can be used to assess the impact of operating characteristics, as well as the effectiveness of different mitigation strategies related to mask wearing, ventilation, etc., on a consistent basis.

**Risk Calculation**

The above model is combined with models simulating the operations of the system to calculate the associated risk for a given OD demand, under different operating practices and mask-wearing behavior, as Figure 2 shows. OD data inferred from smart card transactions (see for example [22]) is input to the subway line simulation.

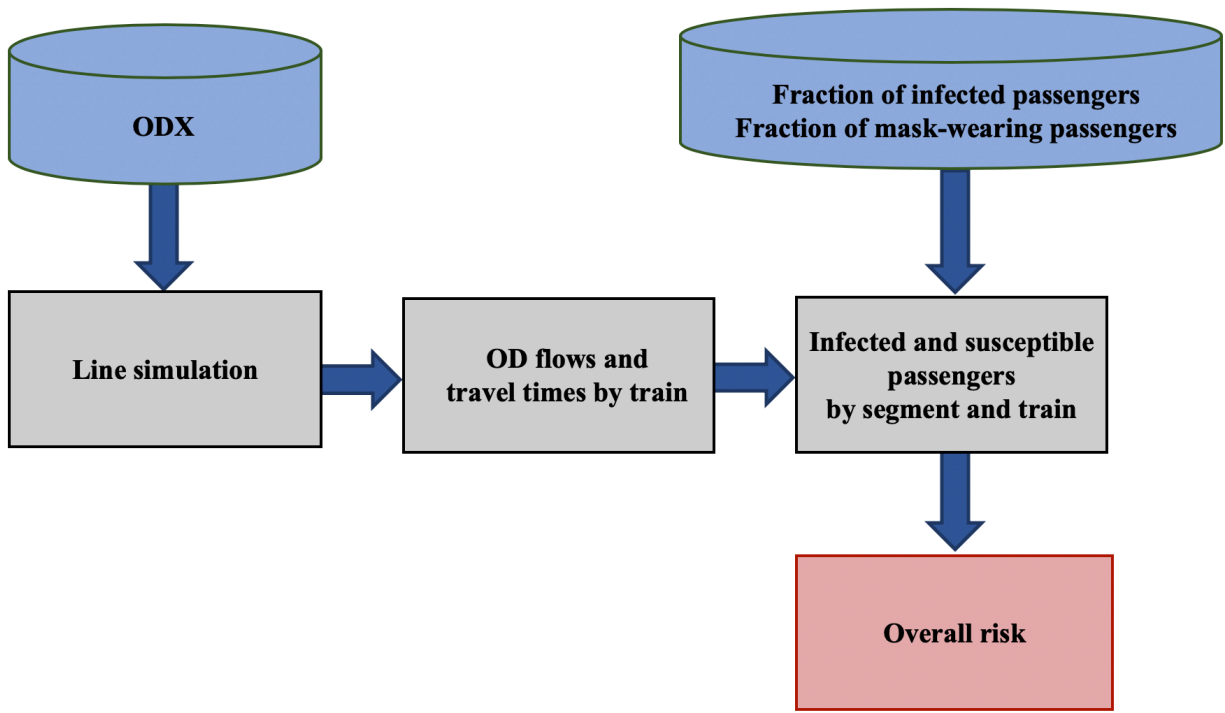

**Figure 2 Simulation based transmission risk assessment for subway systems**

The line simulation model generates the inputs to the risk model at the individual trip level. This includes train loads and passenger journey times between stations. Given the fraction of infectors in the general passenger population (could vary by station) and the fraction of mask-wearing passengers, the model generates the inputs required for the calculation of the infection risk at the train, OD, and system levels for different groups of passengers.

**APPLICATION**

The objective of the case study is to use the methodology proposed in the previous section to evaluate the transmission risk under alternative operating strategies and various assumptions about infectiousness, ventilation, and passenger behavior in terms of mask wearing using a specific heavy rail line. The analysis also explores the effectiveness of different transmission risk mitigation strategies and the sensitivity of transmission risk to different demand levels, different infectiousness levels, and carrier rates in the population. A major subway line in one of the metropolitan areas in the U.S. provides the background for the case study.

To apply the framework described in Figure 2, the model derived in the previous section is combined with an urban rail simulation model (SimMETRO) to assess the risk of transmission. SimMETRO is a microscopic, agent-based stochastic simulation model [23, 24]. It is detailed and designed for heavy rail system performance analysis, operations planning, signaling system evaluation, real-time control strategies evaluation and refinement, and operator-in-the-loop training. It models the network geometry, signaling system, including block design and speed code, providing speed commands for trains to follow. The input includes train schedules and dispatching and rolling stock characteristics. Passenger demand can be modeled at various levels of detail, with dynamic origin-destination (OD) flows (e.g. in 15-minute intervals) at the most detailed level. A detailed dwell time model is utilized to determine a train's dwell times at stations as a function of the number of boarding and alighting passengers.

Various operating strategies can be tested and evaluated in detail by the model under different signal system configurations, passenger demand, schedules, and rolling stock characteristics. Detailed data at the train and passenger levels are generated by the simulation model, such as train runtimes, headways, dwell times, train load at each segment, individual passenger journey times, wait times, and denied boarding due to capacity constraints. The simulation model has been validated in previous studies [24].

**Experimental Design**

Table 1 summarizes the base parameter values for the simulation experiments.

**TABLE 1 Assumptions**

| | Base assumptions |
|---|---|
| 1 | Mask effectiveness in terms of particle penetration rate for exhaling (inhaling) $R_m = 0.5$ ($F_m = 0.5$), as used in [25], to model normal quality face coverings. |
| 2 | $q$ value = 270/hour, back-calculated using the method mentioned in [9] and [15]. Two COVID-19 outbreak cases [10] in public transportation vehicles are used to derive the $q$ value. |
| 3 | Ventilation rate of clean air in a cabin of a subway train $Q$ = 1958 m$^3$/hour [26]. |
| 4 | Breathing rate per person (m$^3$/hour), $p$ = 0.72 m$^3$/hour [27]. |
| 5 | 2% of the passengers carry the virus (infectors), uniformly across all stations. |
| 6 | The air exchange by door-opening at stations is neglected. |
| 7 | Risk due to waiting at stations is not considered. |
| 8 | The time period for the analysis is the PM peak direction, 4:30 to 6:30 PM |
| 9 | The regular frequency of service and schedule is used (no adjustments for changes in demand) |

A number of scenarios were generated using the various factors that impact transmission.
1. *Mask-wearing behavior*. Fraction of population wearing masks set to {0%, 20%, 40%, 60%, 80%, 100%}. Other factors are at their base values (Table 1).
2. *Ventilation rates*. The design vehicle ventilation capacity is set as 100%. Considering the possible degradation of ventilation equipment and possible improvements that increase ventilation rate, the levels of ventilation rate as a percentage of the design ventilation are set to {80%, 100%, 120%, 140%}. Other factors are at their base values (Table 1).
3. *Infectiousness of disease*. Considering different infectiousness levels provides a more comprehensive understanding of the sensitivity of the risk level to assumptions related to the virus transmission rate. We assume $q$ values from the set {170, 190, 230, 270}. Other factors are at their base values (Table 1).
4. *Carrier (infector) rates in the population*. We consider carrier rates at {2%, 4%, 6%, 8%, 10%}. Other factors are at their base values (Table 1). Carrier rate is the same at all stations.

The above scenarios examine the effect of each factor independently. Scenarios with the combined impact of various factors were also evaluated under different demand levels.
1. Mask-wearing behavior (0% - 100%) and ventilation rates (70%-140%).
2. Mask-wearing behavior (0% - 100%) and infectiousness of disease (150-270 quanta/hour).
3. Ventilation rates and infectiousness (70%-140%) of disease (150-270 quanta/hour).

The risk for a given scenario is evaluated for demand levels from 10% to 100% in 10% increments, to capture the demand changes during the various phases of a pandemic.

**Results and Discussion**

*A. Base case*

The base case assumes a pre COVID-19 demand level (OD flows in 15 minute intervals) using the corresponding schedule (during the afternoon peak period). It is assumed that 0% of the passengers wear a mask, there is a 2% carrier rate, the infectiousness level is defined by q = 270 quanta/hour, and the ventilation system operates at its design capacity.

Figure 3 shows the heatmap of the spatial distribution of risk $r_{ij}$ (using equation (18)), in terms of the average number of infections per train trip, for different OD pairs. Stations 6-8 correspond to the downtown area and OD groups which go from the downtown area to the end of the line are having relatively higher numbers of secondary infections. These high-risk areas could be locations for entry control, mask distribution, or stricter social distancing.

We also analyzed the contribution of each station to the overall risk. The results show that the busiest stations are not necessarily the ones that contribute the most. Total entries do not always determine the risk order. For example, in the above network, the highest ranked stations in terms of total entries are stations 8, 9, 10 and 6, but the highest ranked stations in terms of their contribution to the overall risk are 8, 6, 9, 10.

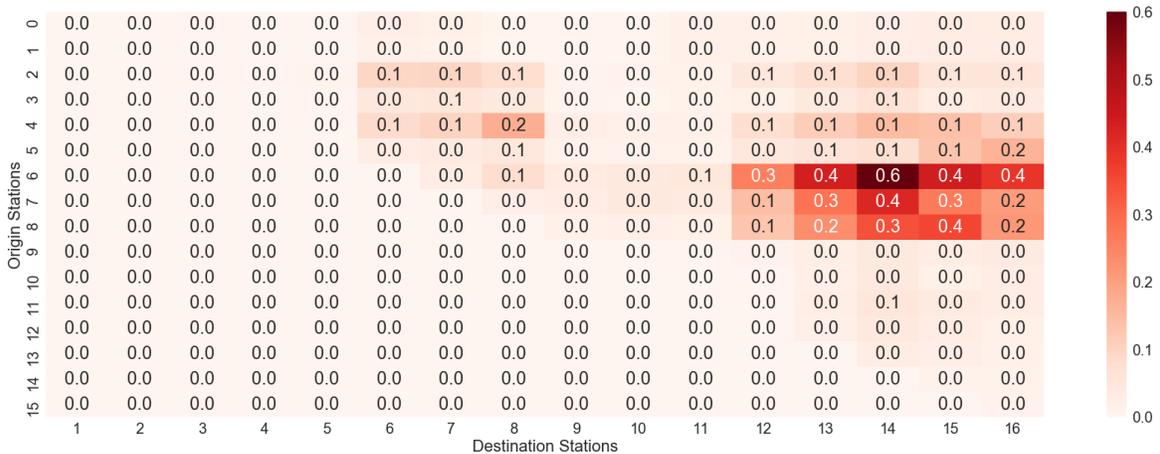

**Figure 3. Expected number of secondary infections per train trip by OD pair**

*B. Impact of mitigation factors*

Figure 4 summarizes the impact on risk at the system level (equation (19)) of various factors at different demand levels (x-axis). The y-axis measures the risk as the ratio to the base case risk.

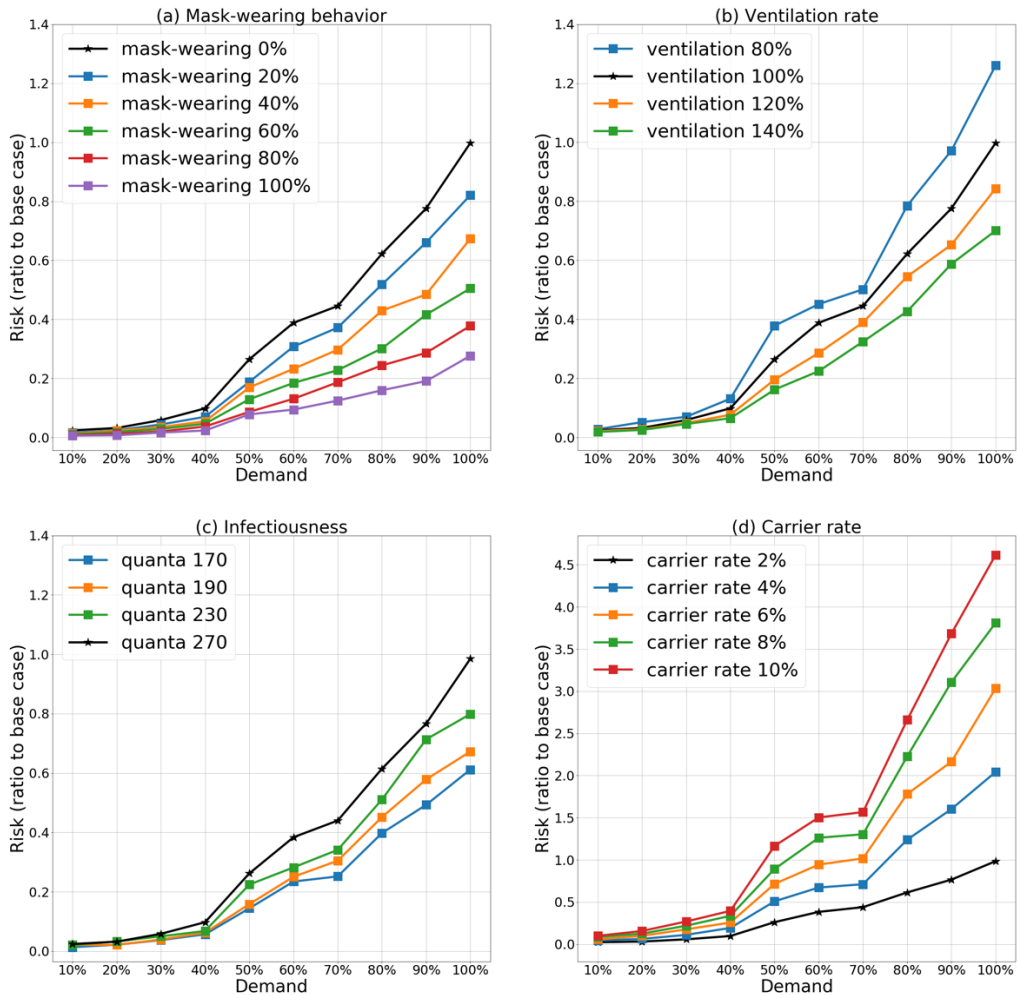

**Figure 4. Transmission risk under different demand levels as a ratio to the base case. (a) Mask-wearing behavior, (b) Ventilation rate, (c) Infectiousness level, (d) Carrier rate**

Figure 4a shows the impact on risk of mask-wearing under various demand levels. When the demand levels are lower than 30% the risk is relatively low, and the differences among the various mask-wearing cases are small. As the demand keeps increasing from 30%, the risk increases more significantly. When passengers wear masks, the risk level increase is not as sensitive compared to the non-masked cases. The sensitivity is lower as the fraction of user wearing masks increases. With 60% of passengers wearing masks, the risk level decrease from the base case is approximately 50% at 100% demand. With 100% of passengers wearing masks, the risk level decrease from the base case is approximately 73%. The mitigation effect of mask-wearing is significant.

Figure 4b shows the impact of ventilation. When the demand levels are lower than 30% the risk is relatively low and the impact of the different ventilation levels is small. With demand above 30%, the risk increases more significantly. Increasing ventilation rates makes the risk level less sensitive to demand increases. Furthermore, decreasing ventilation by 20% increases risk by about 27%. On the other hand, a 20% increase in ventilation reduces risk it by only 15%. Improvements in poor ventilation can have larger impact than improvements where ventilation is satisfactory.

Figure 4c shows the impact of different infectiousness levels. The behavior is similar as in Figures 4a and 4b. Figure 4d shows the impact of different carrier rates. Again, the risk is relatively low and not sensitive to the rate of infectors when demand levels are lower than 30%. At demand levels higher than 30%, the risk increases significantly.

In general, when the demand level exceeds 70%, the risk increases more rapidly, the reason for this are discussed in the next section *D* (operational characteristics).

*C. Combined impact of mitigation factors*

We also explore the impact of combinations of different factors. The scenarios explored in the analysis are summarized in Table 2. In the discussion that follows the risk is expressed as the ratio to the base case system risk (calculated using equation (19)).

**TABLE 2 Scenarios with combinations of different factors.**

|  | Demand levels | Mask-wearing behavior (%) | Ventilation (%) | Infectiousness (quanta/hour) |
|---|---|---|---|---|
| Scenarios 1 | 100% and 60% | 0%-100% | 70%-140% | 270 |
| Scenarios 2 | 100% and 60% | 0%-100% | 100% | 150-270 |
| Scenarios 3 | 100% and 60% | 0% | 70%-140% | 150-270 |

Figure 5a shows the risk heatmap for Scenario 1 at 100% and 60% demand levels, under combinations of ventilation and mask-wearing factors. Lower ventilation and mask-wearing rates result in higher transmission risk. However, differences exist between the two demand levels. At 100% demand, risk levels vary with changes in ventilation rates as well as with changes in mask-wearing popularity. With 70% ventilation rate and 0% mask-wearing rate, the risk is 60% higher than with 100% ventilation. With 100% mask-wearing rate, the risk levels are relatively low regardless of ventilation rates. At 60% demand and mask-wearing rate above 90%, the overall risk levels remain as low as 10% of the base case for all ventilation rates. On the other hand, if no one is wearing a mask, the transmission risk changes significantly as a function of the ventilation rate. The results confirm the importance of mask-wearing in mitigating the transmission risk in public transportation. Even with under-performing ventilation, the transmission risk remains relatively low if more than 90% of the passengers wear masks.

Figure 5b shows the results for Scenario 2. The heatmaps illustrate the risk under combinations of mask-wearing rates and infectiousness levels. With 100% demand, risk levels vary with changes in infectiousness as well as with changes in mask-wearing rates. With infectiousness of 270 quanta per hour, the transmission risk with 0% mask-wearing is more than 2 times higher than with 100% mask-wearing. When mask-wearing is 100%, the risk levels are relatively low for all infectiousness levels. At 60% demand, and mask-wearing rate higher than 80%, the risk levels remain as low as 10% of the base case for all infectiousness values. On the other hand, if the mask-wearing rate is 0%, the transmission risk changes significantly with changes in the infectiousness levels.

Figure 5c summarizes the results for Scenario 3. The risk heatmaps for 100% and 60% demand under different combinations of ventilation rates and infectiousness levels indicate that lower ventilation rates and higher infectiousness levels result, as expected, in higher transmission risk. With 100% demand and infectiousness level of 270 quanta per hour, the transmission risk is 40% higher when ventilation performs at 70% compared to when it operates at normal levels. When the ventilation rate is 140%, the risk levels are relatively low, but the risk is still sensitive to an increase in infectiousness. At 60% demand, even with ventilation rate of 140%, the overall risk level is still relatively sensitive to the increase in infectiousness. The results highlight the effectiveness of ventilation in mitigating the transmission risk in public transportation operations. However, even at high ventilation rates the risk is not negligible.

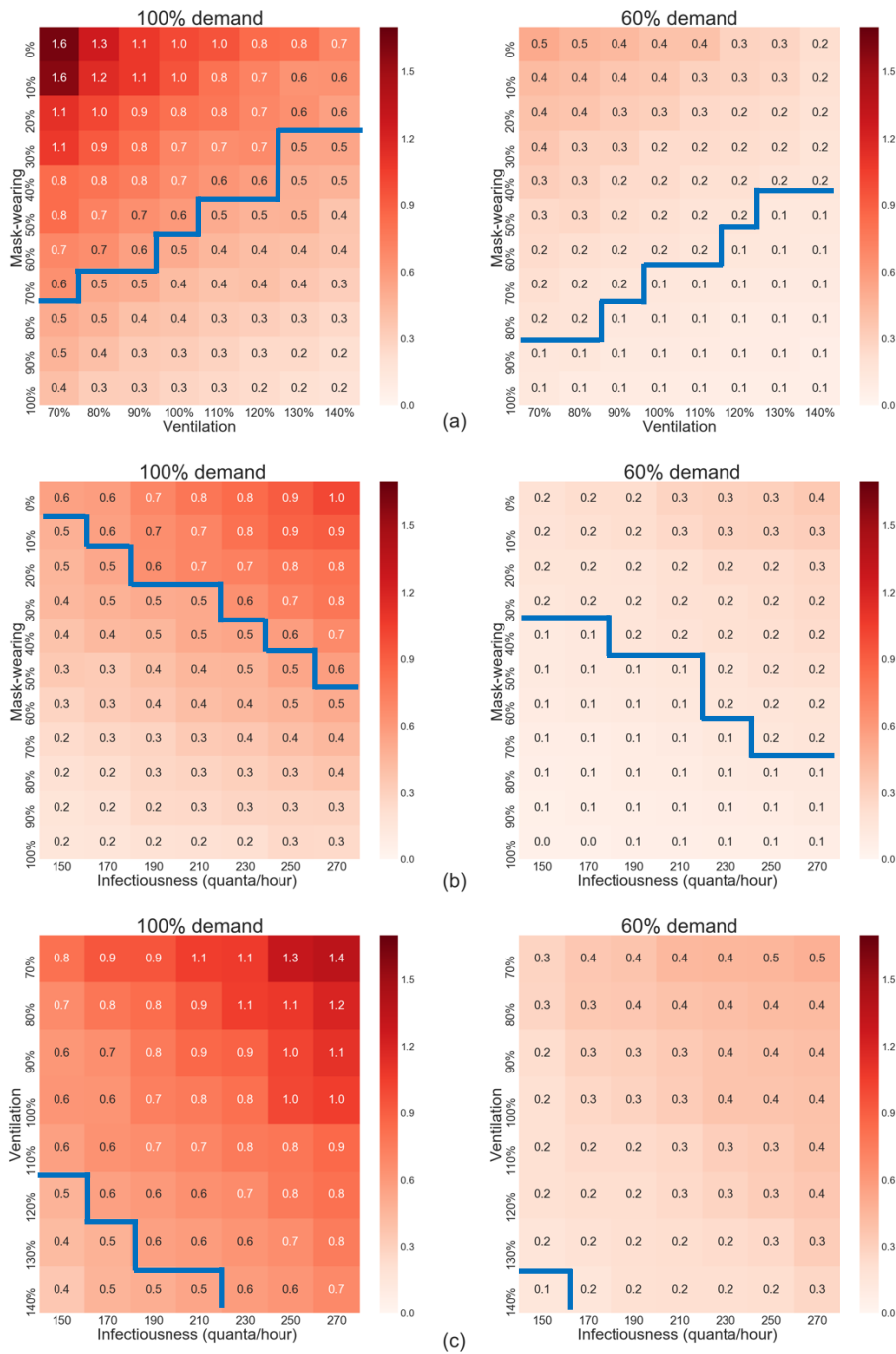

**Figure 5. Risk level for combinations of factors for 100% (left) and 60% demand (right). (a) Mask-wearing and ventilation rates, (b) Mask-wearing rates and infectiousness levels (quanta/hour), (c) Ventilation rates and infectiousness levels (quanta/hour).**

The heatmaps on Figure 5 provide some other insights. The blue line in each group is the boundary below which the risk is more than 50% lower, compared to the base case for 100% demand, and more than 90% lower for 60% demand. Comparing the areas under the blue line, it is clear that mask-wearing is more effective than ventilation (Figure 5b and 5c). Furthermore, reduction of the risk by more than 50% of the base case is attainable with a number of combinations of mask-wearing and ventilation (or reduction of more than 90% in the case of 60% demand).

D. *Impact of operational and service characteristics*

It is worth noticing that in general, in all the scenarios discussed above (for example, section *B*), the risk levels increase more significantly after the demand goes above 70%. This is the result of load and travel time changes as the demand increases. Figure 6a shows the boxplots for runtimes from the terminal station to the peak load station under demand level ranging from 50% to 90%. Figure 6b shows the corresponding boxplots of passenger load per train at the peak load point. The mean and variability of runtimes as well as loads increase significantly as demand increases. The system experiences congestion and delays, and trains are more crowded. The increase in the number of passengers onboard, as well as the increase in runtimes and decrease of reliability, result in the risk increasing at a higher rate for demand levels above 70%.

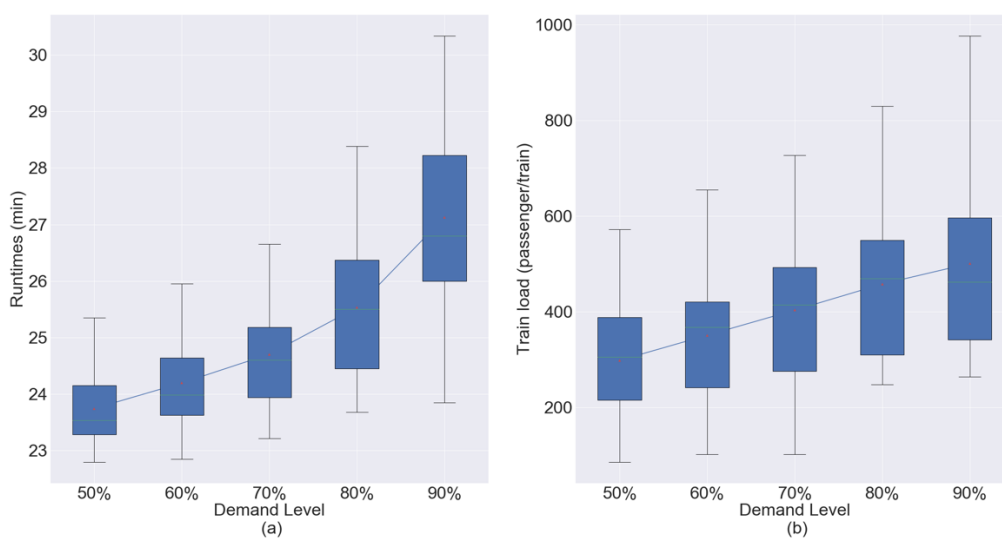

**Figure 6. Performance under different demand levels. (a) Runtimes. (b) Train loads.**

To further explore the impact of service reliability on risk, we compare the risk under two alternative schedules, one with more regular dispatching and uniform scheduled headways during the entire peak period, the other with relatively uneven headways. Both schedules have the same average headway of about 4.3 minutes. The headway coefficient of variation (cv) of the variable schedule is 0.3 and the cv for the regular 0.2. The other settings for the experiment are the same as in Table 1. The demand is 100% and mask-wearing is at 0%.

Figure 7 illustrates the passenger load under the two schedules. The mean loads per train are similar. However, the more uniform schedule results in less variability in train loads. The 90$^{th}$ percentile load for the uniform schedule is about 9% lower than the other alternative.

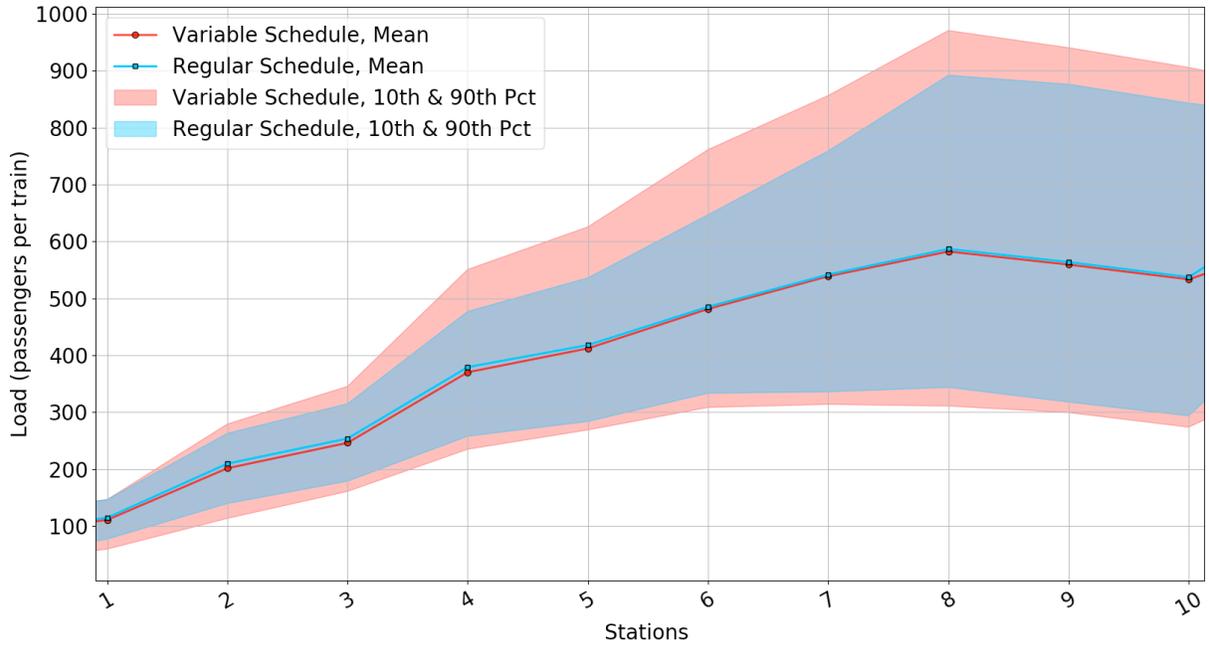

**Figure 7. Passenger load profile under the two alternative schedules**

The variability in loads impacts the distribution of the risk. We use the risk associated with individual trains $r_k$ (equation (17)), the expected infections of a single train trip, to measure the impact. The results show that the more variable schedule results in $r_k$ = 8.2 infections while the regular schedule in $r_k$ = 7.8. The standard deviation for the variable schedule's $r_k$ is 6.0, and the regular schedule 4.8. Schedule regularity can help reduce the risk level, but more importantly it reduces the extreme cases.

Figure 8 shows the histogram of the probability of infection by a random passenger associated with individual trains under the two schedules (equation (20)). With the regular schedule, the mean and standard deviation of this probability is 0.0071 and 0.0046 respectively, while with the more variable schedule, the mean and standard deviation of this probability is 0.0076 and 0.0061 respectively. More importantly the probability of infection of the worst performing train is lower for the regular schedule (0.023 compared to 0.033), a 30% decrease.

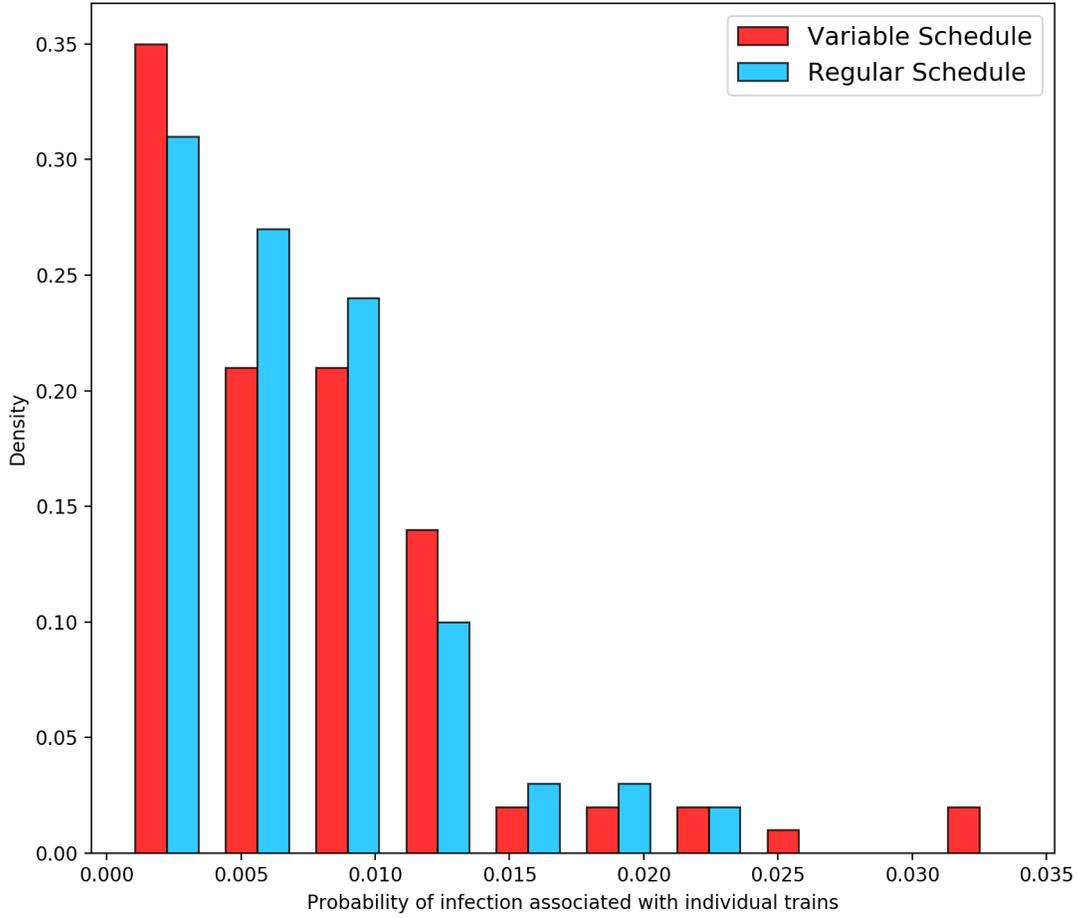

**Figure 8. Histogram of the probability of infection associated with individual trains**

*E. Impact of frequency of service*

While it is important for agencies to adjust their service frequency to meet passenger demand, during a pandemic it is critical to find a balance between operating costs and passenger safety. We assume a demand level at 70% of the base case. Under normal conditions, agencies would adjust the scheduled frequency accordingly. However, under a pandemic, reducing headways may increase infection risk. We hence, explore the impact of frequency of service on risk for a given demand level. The remaining settings for the experiment are the same as Table 1.

Figure 9 shows the infection probability for a random passenger in the system under 7 different headway scenarios (equation (21)). As the headway increases from 4 to 4.5 minutes, the infection probability increases slightly. For headways exceeding 5.5 minutes, the infection probability increases sharply. Increase of headways causes increase in runtimes, dwell times, train load, and the deterioration of service reliability, so the infection probability increases accordingly.

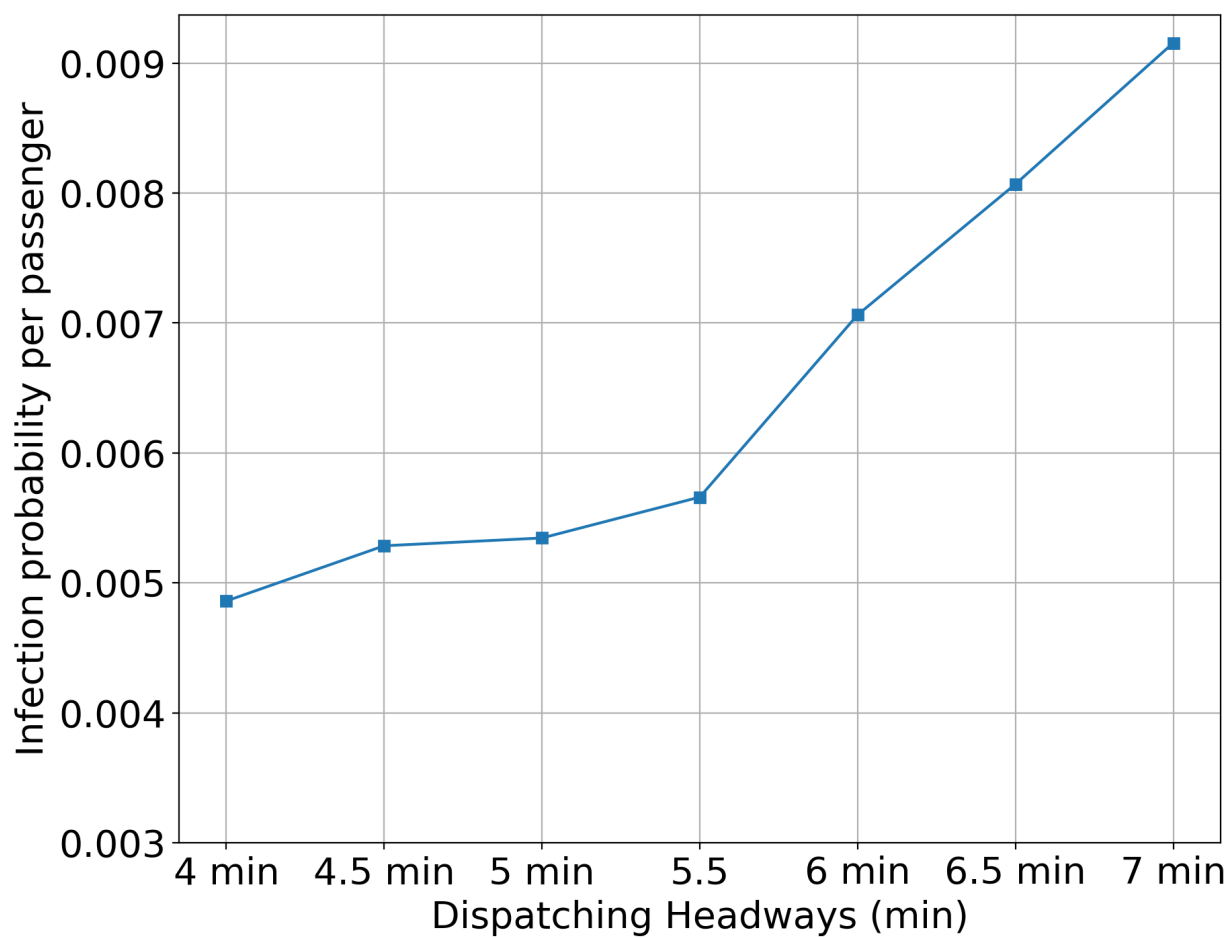

**Figure 9. Risk levels under various dispatching headways**

**CONCLUSION**
The paper explores the transmission risk of airborne viruses through the use of public transportation, in particular subway systems. The Wells-Riley model is utilized to develop an overall risk metric as a function of OD flows, actual operations, and factors such as mask-wearing, ventilation, infectiousness, and virus carrier rate. The model is integrated with a microscopic simulation model of subway operations (SimMETRO) and uses trip by trip data on passenger and train movements as inputs. The risk metric captures the spatio-temporal characteristics of demand and operations and provides means to evaluate the effectiveness of various strategies to reduce the associated risk. The paper explores, using a case study with an actual subway system, the sensitivity of overall risk to various mitigation strategies, including operational strategies, mask-wearing, ventilation rates, as well as factors such as carrier rates in the population, and infectiousness of the virus. The results show that mask-wearing and ventilation are effective mitigation methods under various demand levels, infectiousness, and carrier rates. Mask-wearing is more effective at mitigating risks. Impacts from operations and service frequency are also evaluated, emphasizing the importance of maintaining reliable, regular operations at high frequencies in order to lower transmission risks. The spatial pattern of transmission risk is also explored, identifying high-risk locations.

**AUTHOR CONTRIBUTIONS**
The authors confirm contribution to the paper as follows: study conception and design: J. Zhou, H. N. Koutsopoulos; data collection: J. Zhou; analysis and interpretation of results: J. Zhou, H. N. Koutsopoulos; draft manuscript preparation: J. Zhou, H. N. Koutsopoulos. All authors reviewed the results and approved the final version of the manuscript.